\documentclass[11pt,preprint]{aastex}
\usepackage{apjfonts}

\def\gax{\mathrel{\raise.3ex\hbox{$>$}\mkern-14mu\lower0.6ex\hbox{$\sim$}}}
\def\lax{\mathrel{\raise.3ex\hbox{$<$}\mkern-14mu\lower0.6ex\hbox{$\sim$}}}
\def\gtorder{\mathrel{\raise.3ex\hbox{$>$}\mkern-14mu
             \lower0.6ex\hbox{$\sim$}}}
\def\ltorder{\mathrel{\raise.3ex\hbox{$<$}\mkern-14mu
             \lower0.6ex\hbox{$\sim$}}}

\begin{document}

\title{The Supernova Impostor Impostor SN~1961V: \\ 
   {\it Spitzer} Shows That Zwicky Was Right (Again) }

\author{
   C.~S. Kochanek$^{1,2}$, 
   D.~M. Szczygiel$^{1,2}$,
  K.~Z. Stanek$^{1,2}$
  }

\altaffiltext{1}{Department of Astronomy, The Ohio State University, 140 West 18th Avenue, Columbus OH 43210}
\altaffiltext{2}{Center for Cosmology and AstroParticle Physics, The Ohio State University, 191 W. Woodruff Avenue, Columbus OH 43210}

\begin{abstract}
SN~1961V, one of Zwicky's defining Type~V supernovae (SN),  
was a peculiar transient in NGC~1058 that has variously been categorized
as either a true core collapse SN leaving a black hole (BH) or
neutron star (NS) remnant, or an eruption of a luminous blue variable (LBV)
star.  The former case is suggested by its association with a decaying 
non-thermal radio source, while the latter is suggested by its peculiar
transient light curve and its low initial expansion velocities.  The crucial
difference is that the star survives a transient eruption but not 
an SN. All stars identified
as possible survivors are significantly fainter, $L_{opt} \sim 10^5 L_\odot$,
than the $L_{opt} \simeq 3\times 10^6 L_\odot$ progenitor star at optical
wavelengths.  While this can be explained by dust absorption 
in a shell of material ejected during the transient, the survivor must
then be present as a $L_{IR} \simeq 3 \times 10^6 L_\odot$ mid-infrared source.
{\it Using archival Spitzer observations of the region, we show that such a
luminous mid-IR source is not present.} The brightest source
of dust emission is only $L_{IR} \simeq 10^5 L_\odot$ and does not correspond
to the previously identified candidates for the surviving star.  
The dust cannot be made sufficiently distant 
and cold to avoid detection unless the ejection energy, mass and velocity
scales are those of a SN or greater.  We conclude that SN~1961V was a peculiar, but 
real, supernova.  Its peculiarities are probably due to enhanced mass
loss just prior to the SN, followed by the interactions of the SN blast wave with
this ejecta.  This adds to the evidence that there is a population of SN
progenitors that have major mass loss episodes shortly before core collapse.  
The progenitor is a low metallicity, $\sim 1/3$ solar, high
mass, $M_{ZAMS} \gtorder 80 M_\odot$, star, which means either that BH 
formation can be accompanied by an SN or that surprisingly high mass stars
can form a NS.  We also report on the mid-IR properties of the two other
SN in NGC~1058, SN~1969L and SN~2007gr.  
\end{abstract}

\keywords{supernovae:general, supernovae: individual: SN~1961V, SN~1969L, SN~2007gr}

\section{Introduction}
\label{sec:introduction}

We know that stars both explode, as core-collapse SN, and erupt in
luminous transients that eject mass but do not destroy the star.  In some cases, both types of transients produce
similar, Type~IIn spectra, where the ``n'' indicates that the emission
lines are narrow ($\ltorder 2000$~km/s)  compared to a normal supernova
(\citealt{Schlegel1990}, \citealt{Filippenko1997}).  Type~IIn SNe seem to be cases where the
blast wave is interacting with a dense circumstellar medium created
either by a massive wind or mass ejected in a pre-SN eruption
(see, e.g., \cite{Smith2008}, \cite{Galyam2007} and references
therein).   The mechanism of the eruptions from LBV stars is not
well-understood (see, e.g., \citealt{Humphreys1994}, \citealt{Smith2006}), 
but they eject material at velocities lower than normal SN.  Unfortunately, the luminosities of the 
faintest SN are not well separated from those of the brightest
eruptions, making it difficult to safely classify transients
at the boundary.  These brightest of stellar eruptions are 
frequently referred to as SN ``impostors'' (\citealt{Vandyk2002}).  
Correct classifications are important for understanding
the rates and mechanisms of both processes.  In particular, we note the 
recent debates about the nature of SN~2008S and the 2008 transient 
in NGC~300 (see \cite{Prieto2010} and references therein). 

The most obvious difference between the two cases is that the star
survives only in the eruption scenario.  Thus, there have been attempts to
identify the surviving star for a number of the impostors, with
candidates identified for SN~1954J (\citealt{Smith2001}, \citealt{Vandyk2005}), SN~1961V (see below),
SN~1997bs (\citealt{Vandyk1999}, \citealt{Li2002}), 
and SN~2000ch (\citealt{Wagner2004}, \citealt{Pastorello2010}).
It is probably safe to say that none of these identifications besides SN~2000ch
is certain.  There is, however, a second test.  Most of
the candidate survivors are fainter than the progenitors, and this is 
expected because the surviving star lies inside a shell of 
ejected material that probably forms dust as it cools.  For
the spectacular Galactic example of $\eta$ Carina, $\sim 90\%$ of
the emission is absorbed and reradiated in the mid-IR (see \cite{Humphreys1994}).
Thus, a good test for these identifications is to find 
the mid-IR emission from the survivor and check that it matches the
absorption indicated by the difference between the luminosities
of the progenitor and the survivor.\footnote{There
can be problems in this accounting from binary companions (see \cite{Kochanek2009})
and chance coincidences.}  While some SN may be late time IR sources, they
should evolve more rapidly and are unlikely to show the balance between
progenitor luminosity, optical absorption and mid-IR emission expected for
a surviving star.  While frequently noted, this test never
seems to have been carried out.  We do so here for SN~1961V.

The progenitor of SN~1961V was
(likely) the brightest star in NGC~1058, with $m_{pg} \simeq 18$
in the decades before the transient (\citealt{Bertola1964}, \citealt{Zwicky1964}).
\cite{Utrobin1987} estimated magnitudes in December 1954 of
$B=18.2 \pm 0.1$, $V=17.7\pm0.3$, and $B-V=0.6 \pm0.3$. 
Given a distance of $9.3$~Mpc, the Cepheid distance to fellow
group member NGC~925 (\citealt{Silbermann1996}), and Galactic extinction
of $E(B-V)=0.06$~mag (\citealt{Schlegel1998}), this corresponds to $M_B \simeq -12$,
making the progenitor one of the brightest stars in any galaxy.
Detailed discussions of the light curve are presented in 
\cite{Doggett1985}, \cite{Goodrich1989}, \cite{Humphreys1994},
and \cite{Humphreys1999}, based on the data obtained by
\cite{Zwicky1964}, \cite{Bertola1963}, \cite{Bertola1964}, \cite{Bertola1967},
\cite{Bertola1970}, and \cite{Fesen1985}.  Sometime between
1955 and 1960 the star started to brighten, reaching a plateau of $m_{pg} \simeq 14$
by the summer of 1961 before briefly peaking at $m_{pg} \simeq 12.5$ in 
December 1961.  At this peak, it was brighter than the maximum of the 
Type~IIP SN~1969L (e.g.  \citealt{Ciatti1971})
and comparable to the peak of the Type~Ic SN~2007gr (\citealt{Valenti2008}),
the two other SN in NGC~1058.
It then dropped in brightness, going through a series of extended
plateaus, including a 4 year period from 1963 to 1967 where it
was only moderately fainter than the progenitor, with $m_{pg} \simeq 19$.
After 1968 it had faded below the point of visibility, $m_{pg} \gtorder 22$.
Spectra of the event were also peculiar (\citealt{Branch1971}), with relatively narrow lines
(FWHM $\simeq 2000$~km/s), strong helium emission lines and a constant color, 
resembling an F star, near maximum light (\citealt{Bertola1965}). 
Based on these peculiarities, \cite{Zwicky1964} classified SN~1961V,
along with $\eta$~Carina, as a ``Type V'' supernova, although the
clear presence of hydrogen in the spectra would lead to a ``modern''
classification of Type~IIn or peculiar (\citealt{Branch1985}, \citealt{Filippenko1997}).
  
\cite{Goodrich1989} proposed that the peculiar, extended light curve 
and low velocity spectra would be more easily explained if SN~1961V
was actually an LBV eruption rather than an SN.  They proposed that
the progenitor was a hot ($T_*>45000$~K), luminous ($L_* \simeq 10^{6.4}L_\odot$)
star that was undergoing an S Doradus outburst in the decades prior to the
eruption.  During such an outburst, the star has the same bolometric
luminosity but a far lower photospheric temperature ($T_* \simeq 8000$~K, see
\cite{Humphreys1994}).  This scenario
requires a surviving star, and several candidates have been identified from a
sequence of steadily improving Hubble Space Telescope images by
\cite{Filippenko1995}, \cite{Vandyk2002} and \cite{Chu2004} based
on the accurate optical (\citealt{Klemola1986}) or radio
(\citealt{Branch1985}, \citealt{Cowan1988}, \citealt{Stockdale2001}) positions.
All proposed candidates are significantly fainter
than the progenitor, with $V \simeq 24$~mag.  

The primary counterargument to the LBV eruption hypothesis is that SN~1961V also 
seems to be associated with a fading, non-thermal radio source (\citealt{Branch1985},
\citealt{Cowan1988}, \citealt{Stockdale2001}, \citealt{Chu2004})  that closely resembles the properties of 
other radio SNe and not the fainter, thermal emission of $\eta$ Carina.  There
is no non-thermal radio emission (or even a detection) from the SN impostor/LBV eruption SN~1954J
even though it is almost four times closer and of similar age (\citealt{Eck2002}).  
VLBI observations in 1999 by \cite{Chu2004} also resolved out the radio emission, 
setting a minimum radius for the radio emission of order $4$~mas or about $0.17$~pc.
This would require an expansion velocity of $v \gtorder 4000$~km/s that would be
hard to explain with an eruption.  

While the Spitzer Space Telescope
(SST) was not intended for studies of individual stars at $10$~Mpc,
it should have no difficulty identifying a source with the 
luminosity of the SN~1961V progenitor star in the outskirts of
NGC~1058.  Indeed, \cite{Goodrich1989} note that in the infrared
the source should be
``the brightest point thermal IR source in NGC~1058.''  Here we use archival
SST IRAC (\citealt{Fazio2004}) and MIPS (\citealt{Rieke2004}) data to measure the infrared emission associated    
with SN~1961V.  In \S\ref{sec:data} we discuss the available data,
the astrometry relative to the HST data used to identify candidate
surviving stars, and the resulting estimates and limits on the
mid-IR luminosity.  We model the photometry in \S\ref{sec:models} 
to find that there is insufficient infrared emission for the progenitor 
star to have survived and that SN~1961V must, therefore have been an SN.
We note that \cite{Smith2010} have simultaneously reached this 
conclusion based on the gross differences between SN~1961V and
other supernova impostor candidates and its greater similarity 
to other core-collapse SN.  In \S\ref{sec:othersn} we present the photometry
for SN~1969L and SN~2007gr, the other two SN in NGC~1058.
In \S\ref{sec:discussion} we discuss the consequences of SN~1961V having been an SN.

\begin{figure*}
\centerline{\includegraphics[width=6in]{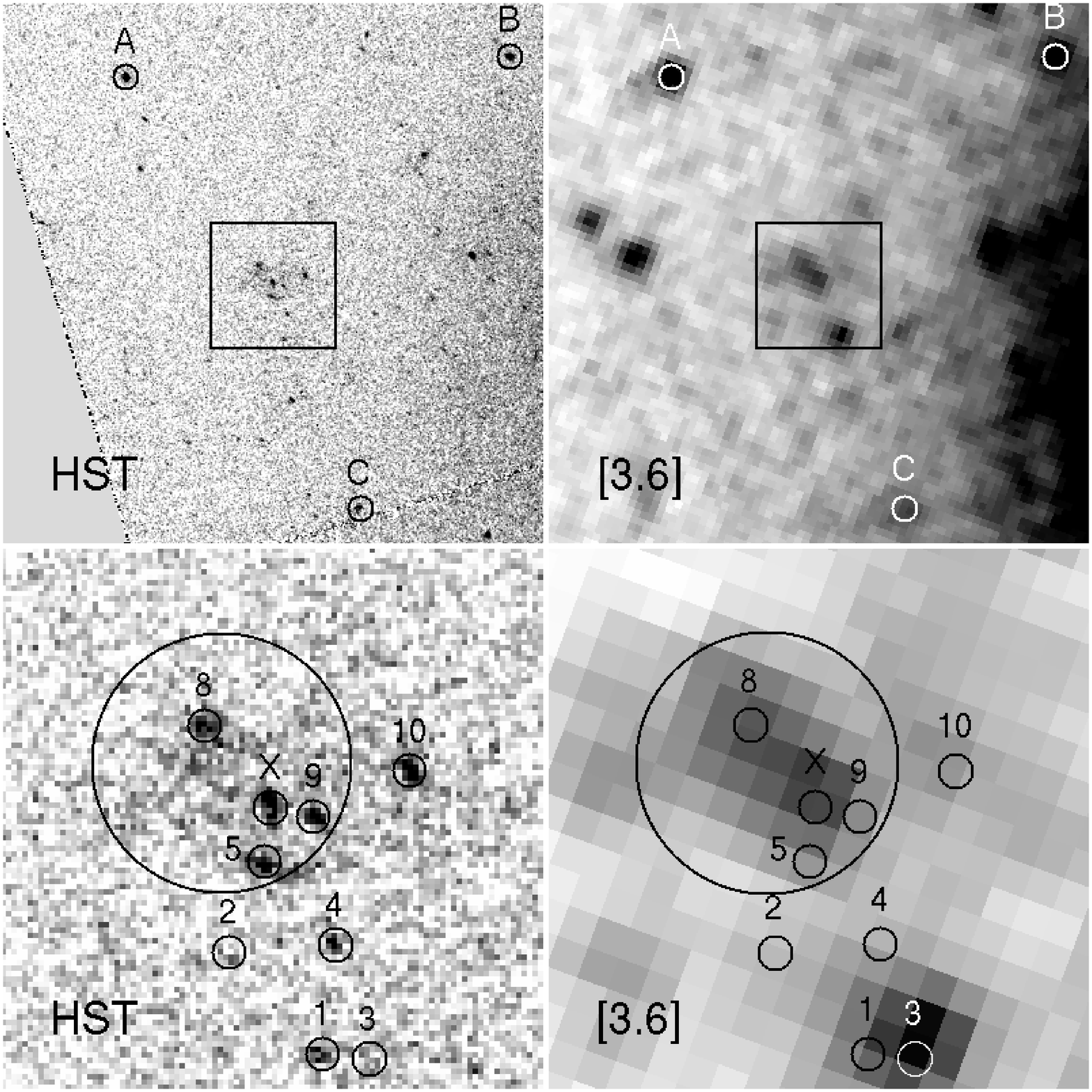}}
\caption{ 
  Astrometric matches between the F606W HST image and the $3.6\mu$m
  reference image.  The top panels show a 45\arcsec\ field of view
  showing three stars (labeled A, B and C and marked by 1\farcs0 radius
  circles) that can be well-matched between the bands.
  The lower panels show a narrower 10\farcs0 region around SN~1961V corresponding to the 
  box in the top panels, where we have labeled the sources following
  \cite{Vandyk2002}. The region labeled X marks the location of stars
  \#6, 7, 9 and 11 from \cite{Vandyk2002} and contains all the candidate
  surviving stars.  Source \#6 is brighter in the F814W image, and
  sources \#3 and 11 are only detected in the F814W image.
  The small circles in the lower panels are 0\farcs3 in radius while
  the large circle corresponds to the 2\farcs4 radius photometry
  apertures shown in Fig.~\ref{fig:photometry}.
  }
\label{fig:astrometry}
\end{figure*}

\begin{figure*}
\centerline{\includegraphics[width=6in]{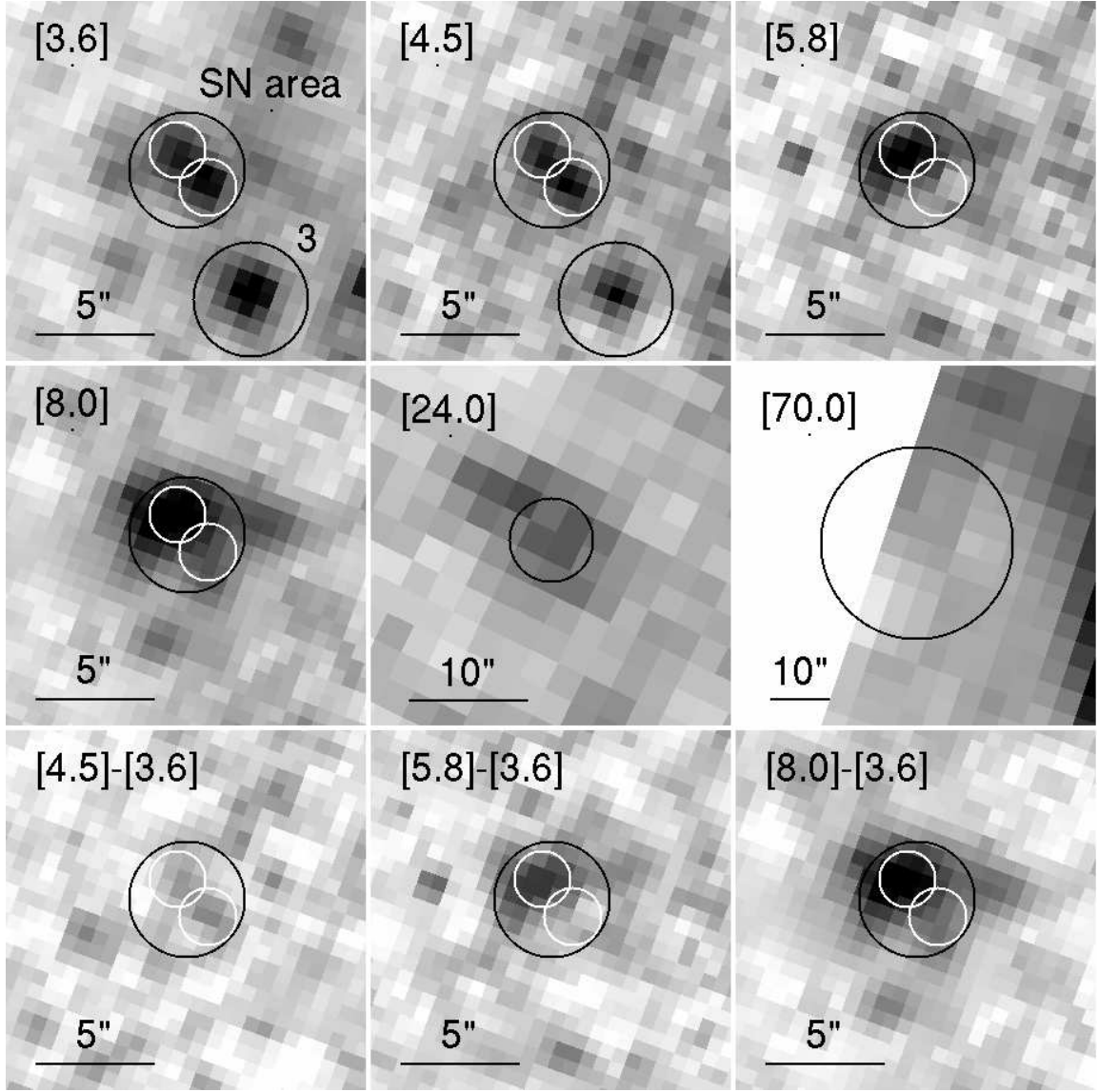}}
\caption{ 
  Mid-IR images and wavelength differenced images of the SN~1961V region.
  The top panels show the $3.6$, $4.5$ and $5.8\mu$m images of the region,
  the middle panels show the $8.0$, $24$ and $70\mu$m images of the region,
  and the lower panels show the $[4.5]-[3.6]$, $[5.8]-[3.6]$ and $[8.0]-[3.6]$ 
  wavelength differenced images.  This removes the flux from normal stars to
  leave only sources of dust and PAH emission.  The panels are 15\farcs0,
  30\farcs0 and 60\farcs0 in size for the IRAC, $24\mu$m and $70\mu$m bands,
  respectively. The large black circles in the IRAC, $24$ and $70\mu$m panels
  have radii of $2\farcs4$, $3\farcs4$ and $16\farcs0$, respectively, and
  correspond to the aperture sizes used for photometry.  The smaller 
  1\farcs2 radius circles mark the positions of star \#8 and the
  region X encompassing the candidate surviving stars.
  We also analyzed the IRAC images of the region using a 3\farcs6 radius
  aperture and DAOPHOT.
  }
\label{fig:photometry}
\end{figure*}

\begin{figure*}
\centerline{\includegraphics[width=6in]{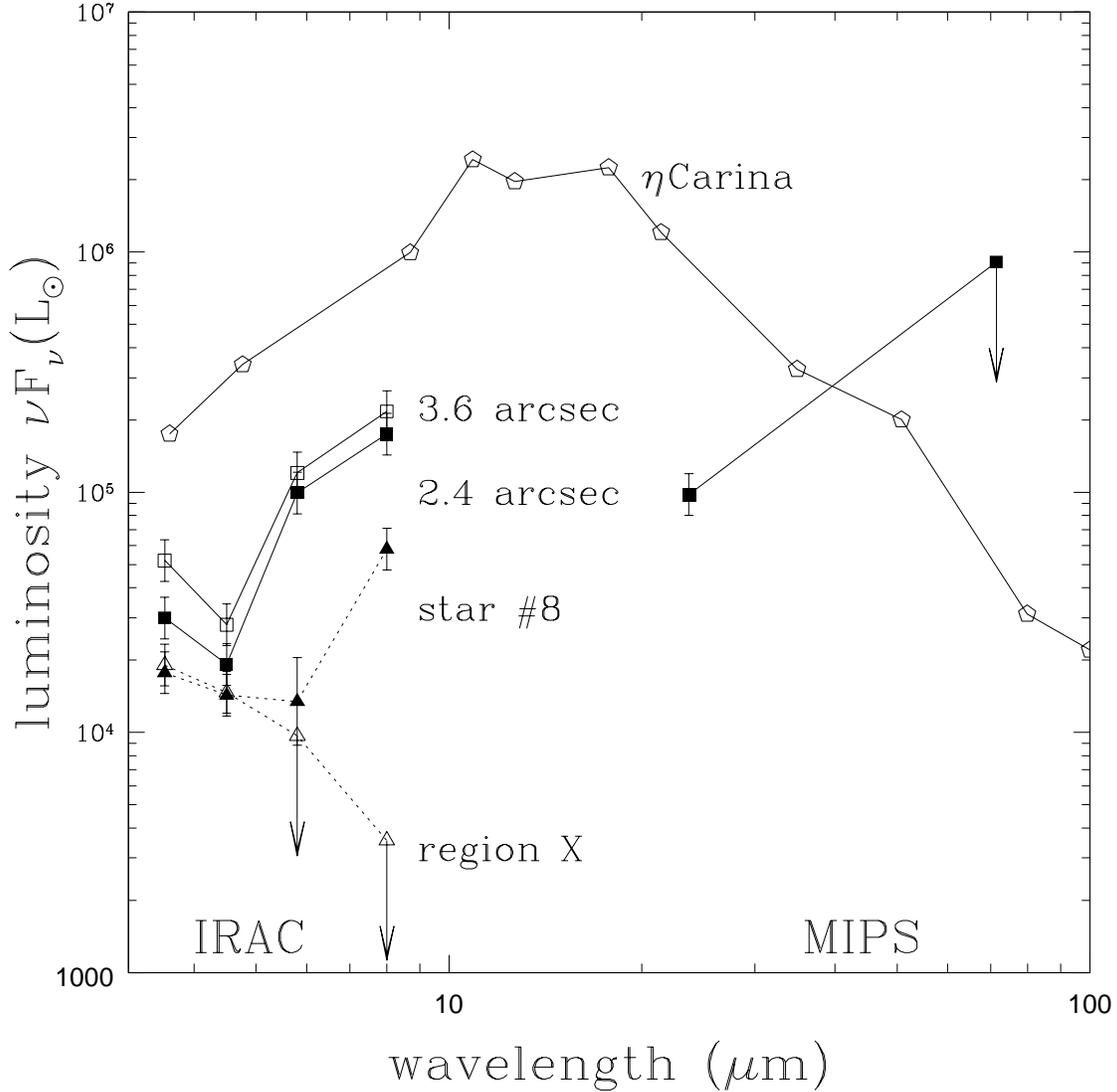}}
\caption{ 
  Mid-IR SEDs of the SN~1961V region.  The total emission is described by the 
  large aperture MIPS fluxes and either the 3\farcs6 (open squares) or 
  2\farcs4 (filled squares) IRAC apertures.  With DAOPHOT we attempt to
  separate the fluxes of star \#8 (filled triangles) and the region X (open triangles)
  that contains all the proposed surviving stars.  For comparison, we show with
  open pentagons the SED of $\eta$~Carina from \cite{Humphreys1994}, which
  roughly has the properties we expect for SN~1961V.  The 2\farcs4 IRAC apertures
  combined with the 24$\mu$m luminosity and the $70\mu$m upper limit will
  be our standard comparison SED.  Note that region X has an SED dropping to
  longer wavelengths,   indicating it is dominated by stellar emission, while 
  star \#8 has an IR excess.
  }
\label{fig:sed}
\end{figure*}

\section{Data and Infrared Luminosity Estimates}
\label{sec:data}

NGC~1058 has been observed twice with IRAC and MIPS, as summarized in
Table~\ref{tab:log}.  The total exposure times are 
$15 \times 30~\hbox{s} = 450$~s for the IRAC bands,
$10~\hbox{s} + 30~\hbox{s} = 40$~s for the MIPS $24\mu$m band and    
$3 \times 3~\hbox{s} = 9$~s for the MIPS $70\mu$m band.    
The SST sensitivity estimates 
for these exposure times are $0.36$, $0.62$, $4.1$, $4.7$,
$25$ and $3400\mu$~Jy, at $3.6$, $4.5$, $5.8$, $8.0$, $24$
and $70\mu$m, respectively.  If we convert these into $3\sigma$
limits on $\nu L_\nu$ in each band at the distance to NGC~1058, they 
correspond to $2400$, $3300$, $17000$, $14000$, $25000$ and
$1.2 \times 10^6 L_\odot$ for the  $3.6$, $4.5$, $5.8$, $8.0$, $24$
and $70\mu$m bands, respectively.  In practice, we would be
confusion limited in the IRAC bands were we trying to reach
these detection limits, but, as \cite{Goodrich1989} noted,
we should have little difficulty finding the expected $> 10^6 L_\odot$ 
mid-infrared source! 

We downloaded the Post-Basic Calibrated Data (PBCD) for these programs
from the Spitzer archive.  These IRAC images are two-times oversampled
and have a pixel scale of 0\farcs60, while the MIPS $24\mu$m and
$70\mu$m images have pixel scales of 2\farcs45 and 4\farcs0
respectively compared to native pixel scales of 2\farcs55 and
5\farcs2 (narrow field of view). 
We aligned and combined the data for each band using the
ISIS (\citealt{Alard1998}, \citealt{Alard2000}) image subtraction package.  We also used ISIS to difference
image between the available epochs, to search for any signs of
variability, and to difference image between wavelengths.  The
latter technique takes advantage of the fact that all ``normal''
stars have the ``same'' mid-IR colors, so normal stars effectively
``vanish'' to leave only the red stars dominated by dust emission
and emission by the interstellar medium (see \citealt{Khan2010}).
This wavelength differencing procedure isolates the relatively 
rare, dusty stars without the crowding from the normal stars.
We also obtained the HST images used by \cite{Vandyk2002} so 
that we could astrometrically match the Spitzer data with the
progenitors discussed by \cite{Filippenko1995}, \cite{Vandyk2002}
and \cite{Chu2004}.  We also examined the more recent images
of the area from October 2007 (Van Dyk/11119) and August 2008 (Li/10877),
but these do not significantly improve on the prior data.

Fig.~\ref{fig:astrometry} shows a wide field and close-up view of the
SN~1961V region in the HST WFPC2 F606W (Illingworth/5446) and $3.6\mu$m 
reference images.  Clearly, with Spitzer's 
resolution we will be unable to obtain photometry for all the individual 
stars identified in the HST image, particularly at the longer
wavelengths.  We see counterparts in the $3.6\mu$m image to star \#8,
the group of stars \#5/6/7/9/11 (which we will refer to as region X)
and star \#3.  Stars \#3 and \#11 are
not visible in the F606W image, but are detected in the F814W
image. The large black circle is $2\farcs4$ in radius and 
represents one of the apertures we used for photometry.

Fig.~\ref{fig:photometry} shows regions around SN~1961V for all 6 Spitzer bands.  We used
black circles to mark the $2\farcs4$ photometric aperture we used to estimate the fluxes on
IRAC images, as well as the $3\farcs5$ and $16\farcs0$ apertures used to measure
fluxes in the $24.0$ and $70.0\mu$m MIPS bands respectively.  For clarity, the alternative $3\farcs6$
aperture used for the IRAC bands is not shown.  We also show the wavelength
differenced images between $3.6\mu$m and the other three IRAC bands.
We see that most of the sources in the $3.6\mu$m image are normal stars,
since they fade away at longer wavelengths and do not appear in
the wavelength differenced images.  There is some dust related emission, much of which
seems to be associated with source \#8, for which we lack an optical color 
because it lay just off the field edge in the F814W and F450W images 
analyzed by \cite{Vandyk2002}, and possibly with source \#10. \cite{Chu2004}
identify star \#7 as the only point-like source of H$\alpha$ emission. 
The complex of sources corresponding to stars \#6, 7, 9 and 11 in \cite{Vandyk2002},
which we have labeled region X in Fig.~\ref{fig:astrometry}, appears to have no
significant excess emission due to dust even though they correspond to all the 
claimed counterparts to SN~1961V (see below).
  
We estimated the fluxes using two procedures.  First, we simply used 
aperture photometry (the IRAF apphot package).  We used signal 
aperture radii (background annuli) of 2\farcs4 (2\farcs4-7\farcs2)
and 3\farcs6 (3\farcs6-8\farcs3) for the IRAC bands, 3\farcs5 (6\farcs0-8\farcs0) 
at $24\mu$m and 16\farcs0 (18\farcs0-39\farcs0) at $70\mu$m.
The background was estimated using the mode of the background pixels
after $2\sigma$ outlier rejection, an approach which should work
reasonably well in crowded regions.  We also compensated for the 
presence of the edge of the $70\mu$m image.  No source was identified
at $70\mu$m, so we estimated a $3\sigma$ upper limit on the flux.
We used the standard Spitzer corrections for these 
apertures.\footnote{http://ssc.spitzer.caltech.edu/irac/iracinstrumenthandbook/ and
http://ssc.spitzer.caltech.edu/mips/mipsinstrumenthandbook/}
For the 2\farcs4 (3\farcs6) IRAC aperture these are 1.213, 1.234, 1.379 and 1.584
(1.124, 1.127, 1.143 and 1.234) for the  $3.6\mu$m, $4.5\mu$m, $5.8\mu$m
and $8.0\mu$m bands, respectively, with uncertainties of order 1--2\%.
For the $24$ and $70\mu$m apertures, they are $2.80$ and $2.07$
and are accurate to about 5\%.  
The resulting flux estimates are presented 
in Tables~\ref{tab:photometry1} and \ref{tab:photometry2}.

While the large aperture photometry provides a conservative upper limit on the
luminosity of any individual source, it is clear that the flux near SN~1961V
can be divided over several sources in the IRAC images.  To better account
for the effects of overlapping PSFs than is possible with aperture photometry,
we also analyzed the region with DAOPHOT (\citealt{Stetson1987}), in particular dividing the
IRAC flux between source \#8 and the complex of sources in region X associated
with the candidate surviving stars.  For
the still lower resolution MIPS images, no attempt was made to divide the
flux over sub-components.  The DAOPHOT results are also presented in
Table~\ref{tab:photometry1}.  Fig.~\ref{fig:sed} compares these estimates
to each other as well as to the SED of $\eta$ Carina from \cite{Humphreys1994}.
The total flux, even in the large apertures, is far less than that of 
$\eta$ Carina, which roughly has the luminosity and SED we
expect for SN~1961V.  The sub-components are then significantly 
less luminous, although the DAOPHOT division into two sources does
not capture all the flux in the aperture, and we again see that while
star \#8 has an IR excess, the region X containing all the proposed
surviving stars seems not to.
We also checked for variability in the IRAC and 24$\mu$m bands, finding
none to limits of roughly 10\%.

We also report photometry for Stars A, B and C in Fig.~\ref{fig:astrometry},
\cite{Vandyk2002} star \#3, SN~1969L (\citealt{Ciatti1971}) and SN~2007gr (\citealt{Crockett2008},
\citealt{Valenti2008}). The 
location of SN~1969L was only covered by some of the images and SN~2007gr 
is only present in the later data, so we only analyzed the relevant
images but followed the same procedures.  We only obtained upper limits
on any flux from SN~1969L, while SN~2007gr was a very bright source.
We discuss the results for these SN in \S\ref{sec:othersn}.

\section{Models}
\label{sec:models}

We model the spectral energy distributions (SED)  using DUSTY (\citealt{Ivezic1997}, \citealt{Ivezic1999},
\citealt{Elitzur2001}).  We assumed a dusty shell with
a density distribution $\propto 1/r^2$ and an outer radius at twice
the distance of the inner, $R_{out}=2R_{in}$.  This assumption has little affect on the
results.  The models are specified by the temperature of the illuminating
black body, the stellar temperature $T_*$, the optical depth of the 
shell $\tau_V$ at V band, and the dust temperature $T_d$ at the
inner edge of the shell.  We tabulated the models for \cite{Draine1984}
graphitic and silicate dusts with the standard size distributions assumed
by DUSTY for stellar temperatures of $T_*=5000$, $7500$, $10000$, 
$15000$, $20000$, $30000$ and $40000$~K, inner edge dust temperatures
from $50$ to $900$~K in steps of $50$~K, and V-band optical depths of $\tau_V=0$ to
$6$ in steps of $0.1$ and $\tau_V = 6$ to $30$ in steps of $0.5$.  
Our approach will be to normalize the models based on the pre-transient 
luminosity and then constrain the optical depth to match the flux of 
the candidate survivors, leaving as the remaining variable the dust
temperature.  

Given the stellar luminosity, the dust temperature is determined by
the shell radius, and the shell radius is closely related to the physics
of the transient.  Since the progenitor must have been 
essentially unobscured pre-transient\footnote{Otherwise, its
already high luminosity would quickly exceed $10^7 L_\odot$
after extinction corrections! Adding additional unrelated 
foreground extinction only strengthens our conclusions because it will 
adjust the progenitor luminosity upwards without contributing
to the mid-IR flux.}, we know
that any obscuring material must have been ejected in 1961.  If
the velocity is restricted by the FWHM of the optical lines, 
roughly $2000$~km/s (e.g. \citealt{Branch1971}), then the current
radius of the material is 
\begin{equation}
    R  \simeq 1.4 \times 10^{17} \left( { v_{ej} \over 1000~\hbox{km/s} }\right)~\hbox{cm}
\end{equation}
where we set the elapsed time to $43$~years (1961 to 2004) and the velocity $v_{ej}$ 
to half of the FWHM.  In our standard models we consider those with inner shell radii
near this value, which is mildly conservative given that the inner edge dominates
the dust emission.  The outer edge is then at twice this distance and so has twice
the expansion velocity. 

The only way to escape our eventual limits is to make the dust so cold that it cannot be 
detected given Spitzer's diminishing sensitivity at longer wavelengths.  For the dust 
temperature to be low, the dust must be distant, and for the simple case of radiative 
equilibrium for dust radiating as a black body, the dust temperature is
\begin{equation}
    T = \left( { L_* \over 16 \pi \sigma R^2 } \right)^{1/4}
      = 142 \left( { L_* \over 3 \times 10^6 L_\odot } \right)^{1/4}
            \left( { 10^{17}~\hbox{cm} \over R }\right)^{1/2}~\hbox{K}
\end{equation} 
(e.g. \cite{Wright1980}) corresponding to an SED peaking near $\lambda = 20\mu$m that will be strongly
constrained by the $24\mu$m data.  DUSTY, with better dust emissivity models,
usually predicts higher inner edge dust temperatures than this simple model
but a similar peak emission wavelength. Lowering the dust temperature to raise
the peak wavelength requires a larger dust radius,  but moving the shell to a 
larger radius requires rapid increases in both the ejected mass and energy.  
The V-band optical depth of the shell is
\begin{equation}
     \tau_{V} = { M_{ej} \kappa_{opt} \over 4 \pi R_{in} R_{out} } 
              \simeq 8 \left( { M_{ej} \over M_\odot }\right)
                       \left( { \kappa_{opt} \over 500~\hbox{cm}^2/\hbox{g} }\right)
                       \left( {  R_{out} \over R_{in} } \right)
                       \left( {  10^{17} \hbox{cm} \over R_{in} } \right)^2
   \label{eqn:opdepth}
\end{equation}
where $\kappa_{opt} \simeq \kappa_{500} = 500$~cm$^2$/g is the optical opacity for a dust
to gas ratio of roughly 1\% (e.g. \citealt{Semenov2003}), $M_{ej}$ is
the ejected mass in which the dust forms, and the shell has a density
profile $\propto 1/r^2$ from $R_{in} < R < R_{out}$.  Equivalently,
the required mass is
\begin{equation}
    M_{ej} = 0.13 \tau_{V} 
                  \left( {  \kappa_{500}  \over \kappa_{opt} }\right)
                  \left( {R_{in} \over R_{out}} \right)
                  \left( {R_{in} \over  10^{17} \hbox{cm} } \right)^2 M_\odot.
\end{equation}
Assuming a thin shell with $R_{in} \simeq R_{out} = R$ for simplicity, the energy of the ejecta, 
\begin{equation}
    E_{ej} = { 1\over 2 } M_{ej} v_{ej}^2 
           = 7 \times 10^{47} 
                \tau_{V}
                \left( {  \kappa_{500} \over \kappa_{opt} }\right)
                \left( {R \over  10^{17} \hbox{cm} } \right)^4~\hbox{ergs},
\end{equation}
increases very rapidly with increasing shell radius because both the velocity and
mass must be larger for larger distances.  The energy required reaches an SN-like 
magnitude of $10^{51}$~ergs for $ R \simeq 6 \times 10^{17} \tau_{V}^{-1/4} $~cm, as
does the velocity and mass.  Making the dust distant enough 
to be cold forces the mass, velocity and energy budgets out of the LBV transient range. 
Phrasing the scaling in terms of the peak wavelength, $M_{ej} \propto \lambda_{peak}^4$ and 
$E_{ej} \propto \lambda_{peak}^8$, further emphasizes the problem with this solution.
Using a thick shell exacerbates these problems since it leads to a larger mass-weighted
radius.

\begin{figure*}
\centerline{\includegraphics[width=6in]{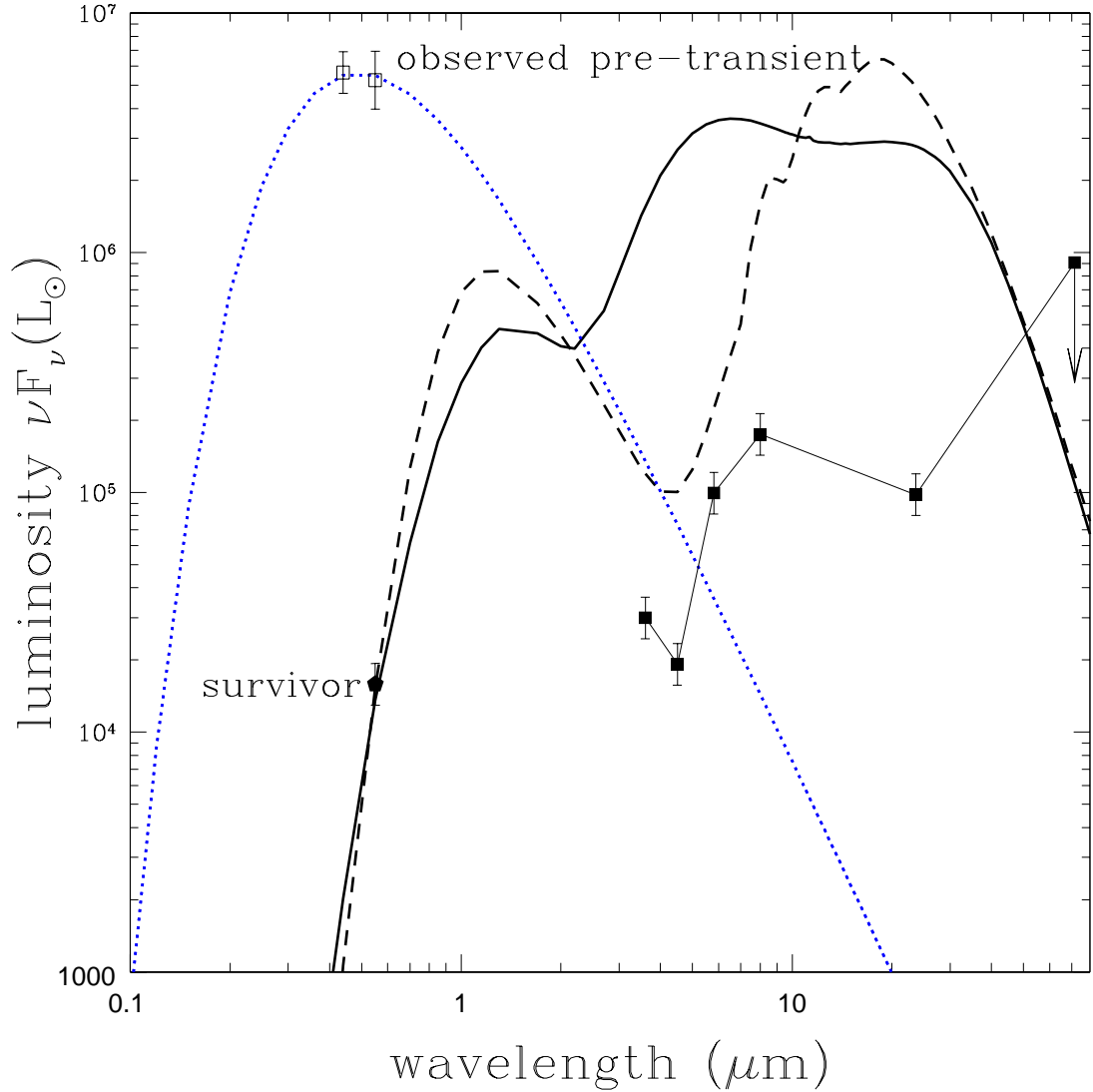}}
\caption{ Models normalized by the pre-outburst magnitudes of \cite{Utrobin1987}.   
  The predicted SEDs for graphitic (silicate) dust shown by the heavy solid (dashed)
  curves lie far above the {\it total} mid-IR emission (filled squares) from the 
  SN~1961V region let alone that of any sub-component (see Fig.~\ref{fig:sed}).  In this case, the progenitor 
  model (light solid curve) is a $T_*=7500$~K black body normalized to match the  
  pre-outburst magnitudes (open squares) from \cite{Utrobin1987}.  The progenitor
  luminosity is $L_* = 10^{6.9} L_\odot$ and it would increase, leading to larger
  discrepancies, if we used a higher or lower stellar temperature. The V band
  optical depths are chosen to match the luminosity corresponding to our 
  generic $V=24$~mag extincted, surviving star (filled pentagon), and the inner
  shell radius is set to be close to $10^{17}$~cm.  
  }
\label{fig:sed1}
\end{figure*}

\begin{figure*}
\centerline{\includegraphics[width=6in]{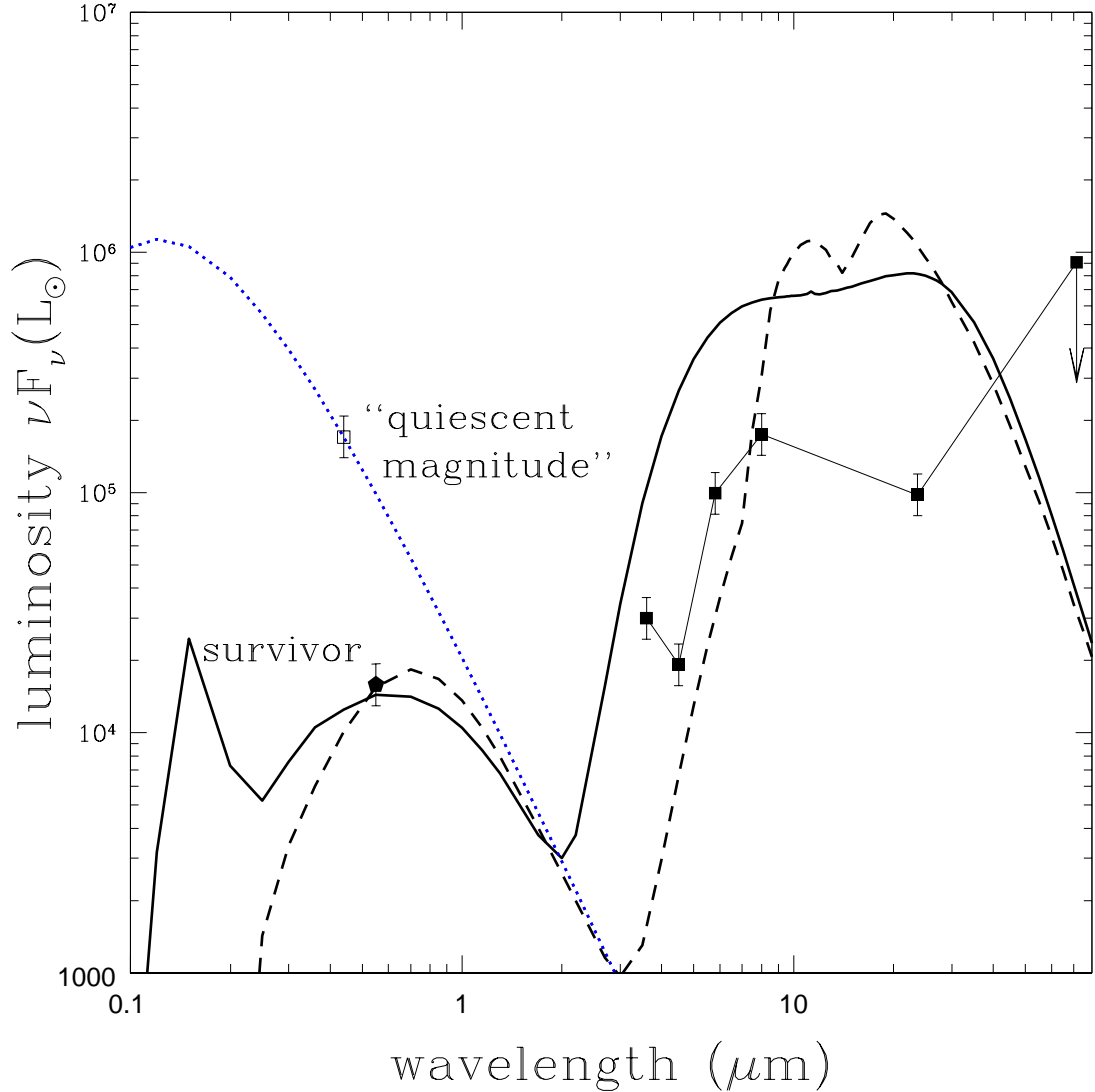}}
\caption{ Models following the \cite{Goodrich1989} scenario.  
  The predicted SEDs for graphitic (silicate) dust shown by the heavy solid (dashed)
  curves still lie well above the {\it total} mid-IR emission (filled squares) from the 
  SN~1961V region let alone that of any sub-component (see Fig.~\ref{fig:sed}).  In this case, the progenitor 
  model (light solid curve) is a $T_*=30000$~K black body normalized to match the
  \cite{Goodrich1989} normalizing magnitude of $B=22$~mag, leading to a stellar 
  luminosity of $L_*=10^{6.2} L_\odot$ that is somewhat low.  The V band
  optical depths are again chosen to match the luminosity corresponding to our 
  generic $V=24$~mag extincted, surviving star (filled pentagon), and the inner
  shell radius is set to be close to $10^{17}$~cm.  For $T_*=40000$~K, $L_*$ would
  double and be closer to matching the luminosity in Fig.~\ref{fig:sed1}, which
  would also double the mid-IR discrepancy.
  }
\label{fig:sed2}
\end{figure*}

\begin{figure*}
\centerline{\includegraphics[width=6in]{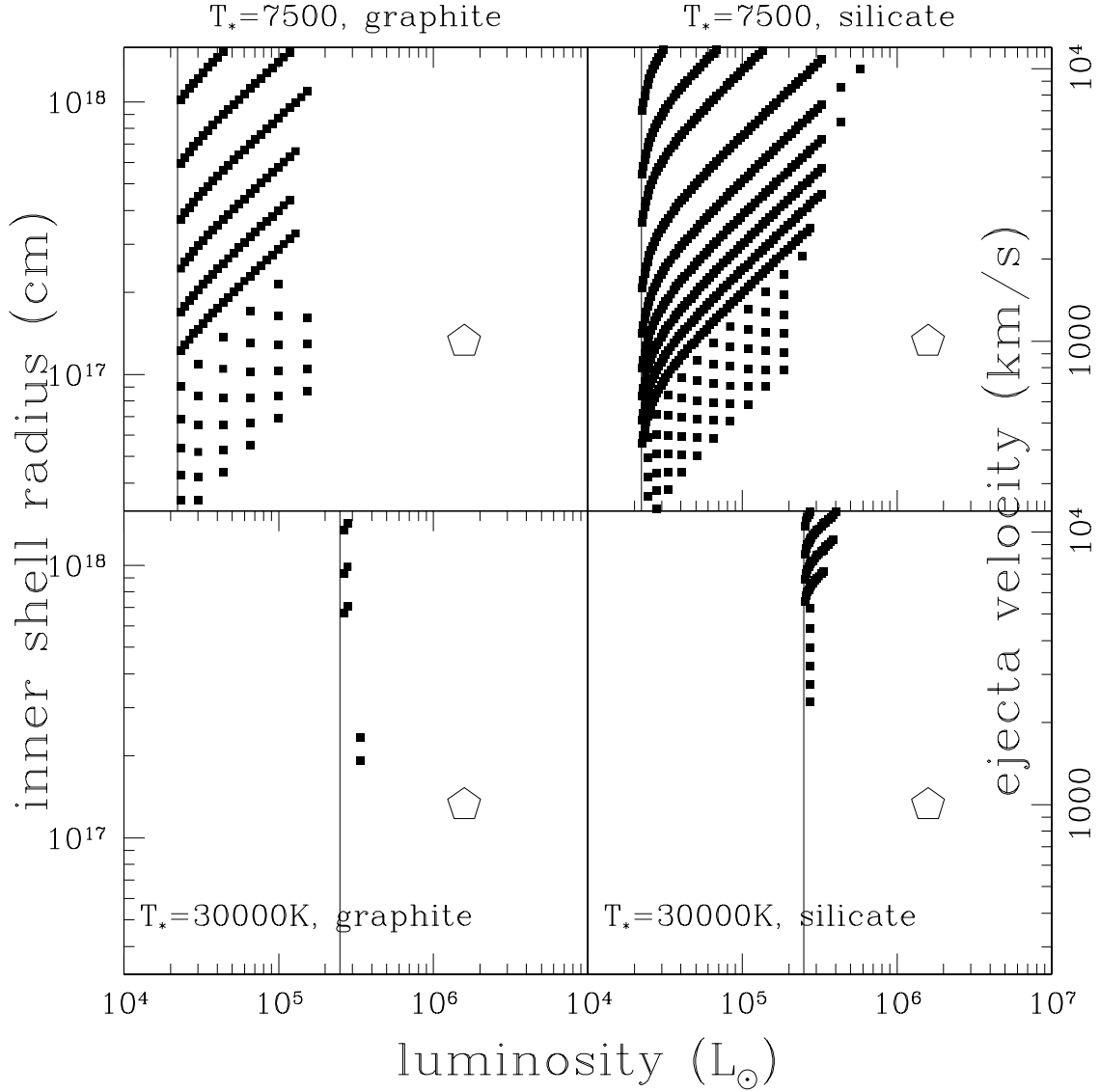}}
\caption{ 
  Stellar luminosities and shell radii (left scale) or ejecta velocities
  (right scale) that fit both the generic $V=24$~mag extincted luminosity of the
  candidate survivors and roughly stay below the upper bound on the
  mid-IR emission ($\chi^2<24$ in Eqn.~\ref{eqn:metric}) for cold 
  ($T_*=7500$~K, top) and hot ($T_*=30000$~K, bottom)
  stars and either graphitic (left) or silicate (right) dust. 
  The vertical lines indicate the minimum ($\tau_V=0$) luminosity consistent
  with a $V=24$~mag survivor.
  The open pentagon marks the solution with the properties typically
  associated with LBV transient hypothesis at $L_*=10^{6.2}L_\odot$ and $v_{ej}=1000$~km/s.
  }
\label{fig:limits}
\end{figure*}

\begin{figure*}
\centerline{\includegraphics[width=6in]{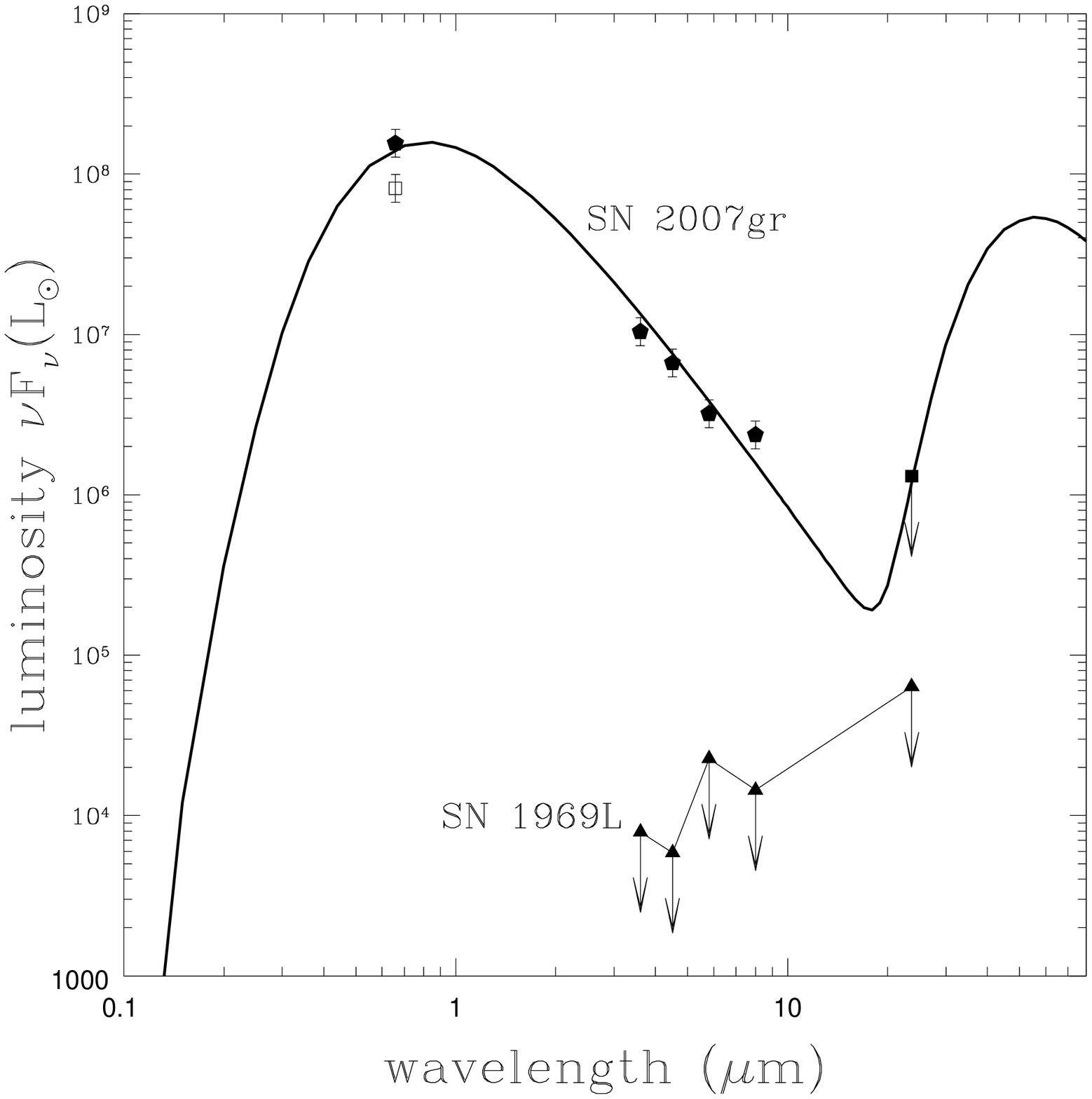}}
\caption{
  Limits on the mid-IR SED of SN~1969L (filled triangles) and the observed
  SED of SN~2007gr (filled squares).  The filled R-band point from 
  \cite{Valenti2008} corresponds to the epoch ($+17$ days) of the IRAC observations 
  while the open point corresponds to the epoch ($+31$ days) of the MIPS observations.
  The heavy solid curve is a $5000$~K black body with luminosity $10^{8.4}L_\odot$.
  }
\label{fig:2007gr}
\end{figure*}

In order to understand the mid-infrared limits we must first choose which HST star
to call the survivor (see Fig.~\ref{fig:astrometry}). \cite{Filippenko1995} propose star \#6, with B and V
magnitudes of $24.82\pm0.25$ and $24.50\pm0.16$~mag.  
\cite{Vandyk2002} propose star \#11, which they estimate to have
an I magnitude of $24.3 \pm0.22$, $B-I > 1$ and $V-I > 1.1$~mag.
\cite{Chu2004} propose star \#7 from \cite{Vandyk2002}, with  B, V and I magnitudes of 
$24.04\pm0.14$, $23.85\pm0.14$  and $23.83\pm0.14$, respectively.  \cite{Chu2004}
prefer this identification because (1) it appears to be spatially coincident
with the radio source, (2) it appears to be the only H$\alpha$ source, and
(3) its H$\alpha$ line has broad wings (at least $\pm 550$~km/s, limited by
the noise in the spectrum).  It does not, however, have the forbidden lines
([OI] $\lambda 6300$\AA, [OIII] $\lambda 4959/5007$\AA) expected from a remnant, suggesting the H$\alpha$
emission is stellar.  It is also too blue to be well-modeled as an extincted
(hot) star, as noted by \cite{Chu2004}.   Of the \cite{Vandyk2002} candidates
closest to the preferred positions, \#5 and \#9 also have the wrong spectral
slopes, while \#6 and \#11 are easily fit.  Star \#8, which has not been
proposed as a candidate because it is too distant, 
is the only nearby source with significant dust emission (see Figs.~\ref{fig:photometry} 
and \ref{fig:sed}). 

In practice, it matters little which star we use for the survivor.  
They are all faint compared to the
progenitor star and so must be heavily extincted in the optical.  Once most of the 
optical/UV flux must be absorbed, it matters little for the expected
mid-IR luminosities whether it is 90\% or 99\%.  Similarly, the 
spectral shape of the candidate matters little, since the optical depth
is principally determined by the magnitude difference between the survivor
and the progenitor rather than the color.  Thus, for simplicity
we will simply give the survivor the typical magnitude of the candidates,
$V=24$~mag, and ignore the colors.   

We have two possible choices for the intrinsic properties of the star. First,
we can simply normalize it using the pre-transient luminosities.  Alternatively,
we can follow \cite{Goodrich1989} and assume that the star was in an S Doradus
phase just before the transient and has now returned to its hotter, but similar
luminosity quiescent state.  If we set the stellar temperature to $T_*=7500$~K
and normalize it by the \cite{Utrobin1987} magnitudes, we get a luminosity of
$L_* \simeq 10^{6.9} L_\odot$.  This is higher than the \cite{Goodrich1989}
proposal of $10^{6.4} L_\odot$, but matches their proposed S Doradus outburst
temperature and agrees tolerably well with the $B-V$ color from \cite{Utrobin1987}. 
Both raising and lowering the assumed temperature increases the luminosity 
because for $T_*=7500$~K the peak of the SED lies near the normalizing 
photometric bands.  In fact, with any significant change in the temperature,
the luminosity becomes impossibly large ($L_* > 10^7 L_\odot$).    

Fig.~\ref{fig:sed1} shows the resulting models.  Dust optical depths of
$\tau_V=7$ and $11.5$ for graphitic and silicate dusts lead to enough
extinction to make the optical flux consistent with a $V=24$~mag flux
for the surviving star.  Choosing inner edge dust temperatures of $T_d=500$
and $300$~K corresponds to putting the inner edge for the dusty shell 
at $R \simeq 1.3 \times 10^{17}$~cm.  The outer edge dust temperatures are
roughly $140$~K.  Using a thin shell, with $R_{out}=1.2 R_{in}$
instead of $2 R_{in}$ leads to no significant changes.  It is immediately
apparent that the predicted mid-IR emission is grossly discrepant with
the constraints even when we compare the model to the integrated emission
from the region (defined here by the 2\farcs4 IRAC fluxes and the MIPS
aperture fluxes) without any division of the emission over the multiple
sources within it (see Fig.\ref{fig:sed}).  
The cool stellar temperature exacerbates the problem
because much of the near-IR emission is little affected by extinction.

The alternate hypothesis, that the star has reverted to a quiescent
hot state (\citealt{Goodrich1989}), changes things little.  Here we
assume that the surviving star has now left its S Doradus phase, and 
again has a high photospheric temperature with a quiescent magnitude that
is $4$~mag fainter than before the eruption, $B \simeq 22$~mag.  For black
bodies with $T_*=30000$~K and $40000$~K, this implies luminosities 
of $L_* = 10^{6.2} L_\odot$ and $L_* = 10^{6.5} L_\odot$ that are
significantly below that implied by the pre-eruption luminosities, as
also noted by \cite{Humphreys1999}. 
Fig.~\ref{fig:sed2} presents the $T_*=30000$~K  models.  The visual optical depths
are now much smaller ($\tau_V=2.5$ and $4.5$ for graphitic and silicate dusts)
because most of the flux is in the UV where the dust opacities are higher.
Choosing inner edge dust temperatures of $T_d=400$
and $300$~K corresponds to putting the inner edge for the dusty shell 
at roughly $R \simeq 1.5 \times 10^{17}$~cm.  The outer edge dust temperatures are
roughly $100$~K.  Again, using a thin shell leads to no significant changes.
The discrepancies are smaller here, partly because the stellar luminosity
is a factor of 5 lower than in the models of Fig.~\ref{fig:sed1}, and
partly because the star has little near-IR luminosity compared to the
cooler model.  Nonetheless, the predicted mid-IR fluxes are still
much higher than allowed by the observations. Raising the stellar 
temperature to $T_*=40000$~K in order to better match the pre-transient 
luminosity or the \cite{Goodrich1989} models makes the problem worse by
a factor of two. 

In the end, the two most important variables are the intrinsic luminosity
and the radius of the dust shell, or equivalently the ejection velocity
of the material.  We can explore these models by normalizing our DUSTY
models to fit the generic $V=24$~mag of the obscured progenitor star 
and then keeping only the models consistent with the total luminosity
of the region, $L_{obs}(\lambda_i)$, again defined by the 2\farcs4 aperture
IRAC luminosities and the MIPS aperture luminosities and limits.
The shell radius, or equivalently the expansion velocity, is the primary 
secondary variable.  We can quantify the consistency using the mid-IR
luminosities for the region as luminosity limits based on the metric 
\begin{equation}
     \chi^2 = \sum_i \left( { L_{mod} (\lambda_i) / L_{obs} (\lambda_i) } \right)^2
     \label{eqn:metric}
\end{equation}
where $\lambda_i$ corresponds to the 6 Spitzer bands, $L_{obs}(\lambda_i)$
is the Spitzer luminosity limit and $L_{mod}(\lambda_i)$ is the luminosity 
predicted by the model.  An SED passing exactly through these 6 values
or limits would have $\chi^2=6$,
while those that are too bright will have higher values and those that are
fainter will have lower values.  At least for the IRAC bands, we know that
these large aperture luminosities are subdivided over multiple sources and 
so are upper limits even if any of the source identifications are correct.
In Fig.~\ref{fig:limits} we show all cases with $\chi^2 < 24$, which corresponds
to the typical model luminosity being twice the observed upper limit, 
for stellar temperatures of $T_*=7500$ and $30000$~K and for graphitic
and silicate dusts.  As the secondary parameter we use the radius of 
the inner edge of the shell, or equivalently the expansion velocity 
of the shell.  We only show cases with $\tau_V \geq 0.1 $.

If the star is hot, $T_*=30000$~K, then the stellar luminosity has to be
at least $L_* > 10^{5.2} L_\odot$ in order to fit the optical luminosity
of the surviving star.  This is already high enough to make it very 
difficult to fall below the upper limits on the mid-IR luminosities in
the presence of any dust. 
A few silicate models work by putting the shell so far out that the 
dust is cold enough to avoid all but the weak $70\mu$m
limit or by making the optical depth significantly less than unity.  
More solutions are possible when the star is $T_*=7500$~K,
because the star is intrinsically less luminous for the same V
band normalization.  We see the expected trend allowing more
luminous stars for more distant and colder shells.  There are
no solutions in the regime required by the LBV eruption hypothesis.

We see no signs of variability in the mid-IR at the level of about
10\% of the IRAC fluxes, although we would not
expect to given the limited time baseline.  While the existing data
is not of high enough quality to make the test, we note that the 
optical variability should be significant given the parameters of
our models.  As the optical depth drops, the source should become
steadily brighter, as is observed for $\eta$ Carina (e.g. \citealt{Humphreys1994},
\citealt{Humphreys1999}).  If we normalize the optical depth to $\tau_1$
and time $t_1$, the optical depth scales as $\tau = \tau_1(t_1/t)^2$
(Eqn.~\ref{eqn:opdepth}),
although the expected magnitude does not simply scale with $\tau$ 
because it includes both scattering and absorption.  If we take 
$\tau_V = 4.5$ from the silicate models for a hot star, then
a $V=24.0$~mag progenitor in 2004, then it should have been
$25.7$~mag in 1995, and should be $23.4$~mag in 2010,
$23.1$~mag in 2015, and $22.9$~mag in 2020.  In the cool
star models the evolution is even more dramatic because
of the higher optical depths.  As \cite{Chu2004} noted, there is 
no sign of optical variability in the published data. 

\section{SN~1969L and SN~2007gr}
\label{sec:othersn}

Since we were analyzing the Spitzer data already, we measured the fluxes associated
with the other two SN in NGC~1058, SN~1969L and SN~2007gr, reporting their fluxes
in Tables~\ref{tab:photometry1} and \ref{tab:photometry2} and presenting their SEDs in Fig~\ref{fig:2007gr}.  
SN~1969L lay outside the $70\mu$m image, and the image was taken before SN~2007gr, so we
have no information on their $70\mu$m fluxes. We detected no flux 
above background at the location of SN~1969L (\citealt{Ciatti1971}), at limits on
$\nu L_\nu$ of $10^4$ to $10^5 L_\odot$. Here we used the $3\sigma$ upper bounds
from the 3\farcs6 radius IRAC aperture and the normal $24\mu$m aperture.
The observations of Type~Ic SN~2007gr (\citealt{Crockett2008}, \citealt{Valenti2008})
were taken 17 (IRAC) and 31 (MIPS) days after the R-band peak. The mid-IR luminosities
are comparable to what was expected for SN~1961V, with luminosities $\nu L_\nu$ of
$10^{7.0}$, $10^{6.8}$, $10^{6.5}$, $10^{6.4}$ and $10^{6.1}$ for the $3.6$, $4.5$,
$5.8$, $8.0$ and $24\mu$m bands respectively.  The SED, as shown in Fig.~\ref{fig:2007gr},
is falling rapidly in the mid-IR, indicating that the emission is not dominated
by a cool dust echo.  We estimated R-band magnitudes at the two epochs of $13.6$
and $14.3$~mag based on the light curve in \cite{Valenti2008}.  The combined
optical and mid-IR SED is well fit as a 5000~K black body with a luminosity
of $10^{8.4} L_\odot$.  If we look at the wavelength differenced
$4.5\mu$m image from before the SN, there does appear to be excess emission,
consistent with the presence of the stars having K band excesses in \cite{Crockett2008},
but with the resolution of Spitzer and the presence of bright stars just North
and South of the site, we cannot say more.

\section{Discussion}
\label{sec:discussion}

The basic conundrum is simple.  All possible surviving stars are far
fainter than the progenitor in the optical.  This requires significant visual optical depths
so that most of the surviving star's luminosity is re-radiated in the
mid-IR.  However, this opacity must be supplied by the material ejected
during the transient, and dusty shells ejected at the low velocities 
of the LBV hypothesis and irradiated by a surviving star lead to mid-IR 
luminosities in gross conflict with the observational limits.  The 
discrepancy is not a subtle problem, but a disagreement of an order
of magnitude or more.  The limits could be evaded by pushing the shell 
so far outward in radius that the dust temperature is too low to have
significant emission in the $3.6$ to $24\mu$m range, but this solution
requires shell masses, velocities and energies that are extreme
even for a true SN.  The simplest solution to these problems is that 
SN~1961V was in fact a supernova and there is no surviving star.

To escape the conclusion that SN~1961V was a supernova requires that
an assumption about the properties of the star or its surrounding dust
is greatly in error. We can enumerate three  possibilities for
changes in the stellar properties: 
(1) the system was (bolometrically) super-luminous for $\sim 3$ 
decades prior to the transient; (2) the surviving star has been
(bolometrically) sub-luminous for the $\sim 5$ decades after the
transient; and (3) there is no dust and the star now has a very
high photospheric temperature, $T_* \simeq 70000$ to $100000$~K, 
so that the faintness of the survivor is entirely due to bolometric 
corrections.  In (1) and (2), the change in bolometric luminosity
must be an order of magnitude or more.  In case (3), such a radical 
increase in photospheric temperature would have be accompanied by significant mass loss which would be
hard to reconcile with the requirement for no dust. Deeper ultraviolet 
observations than the available GALEX data would constrain this possibility,   
The remaining possibility is that the dust covering
fraction is very small, less than 10\%, with our line of sight coincidentally 
passing through one of the optically thick patches.  The mid-IR emission 
is then reduced by the covering fraction.  In this scenario we should 
still see the steady optical brightening created by an expanding shell. 
None of these possibilities seems terribly attractive or plausible.

If SN~1961V was a supernova, then it becomes one of the rare SNe
with observations of its progenitor star, and the only one with
a relatively detailed pre-explosion light curve (see \citealt{Smartt2009}).
We know it was very luminous, $L_* \sim 10^{6.3} L_\odot$, and
likely hot in quiescence following the arguments of \cite{Goodrich1989}.
Based on the emission line ratios reported by \cite{Goodrich1989}
for the Western HII region (the Eastern HII region overlaps the region 
with SN~1961V and shows evidence for contamination by a SN remnant),
we estimate an oxygen abundance of approximately 8.3 following \cite{Kewley2008}
or approximately $\sim 1/3$ Solar and similar to the metallicity of the LMC.  This
local measurement is a little lower than estimates of $\sim 1/2$ Solar from the
metallicity gradient measurements by \cite{Ferguson1998}.  If we select stars at the end points
of the Padua (\citealt{Marigo2008}) isochrones with $10^{6.1} L_\odot < L_* < 10^{6.5}$
and $20000~\hbox{K} < T_* < 40000~\hbox{K}$, they correspond
to very massive stars, with $M_{ZAMS} > 80 M_\odot$ for all
metallicities from LMC to Solar.

These properties are very similar to those of the only other high
mass progenitor to be identified, that of SN~2005gl (\citealt{Galyam2007},
\citealt{Galyam2009}).   For their measurements
and parameters (a progenitor with $V=20.04 \pm0.15$~mag at 66~Mpc with
$E(B-V) \simeq 0.07$), this progenitor had a luminosity similar
to that of SN~1961V, with $L_* \simeq 10^{6.7} L_\odot$ for $T_*=20000$~K.
Like SN~1961V it was a Type~IIn, and it had a comparable peak luminosity,
near $M_V \sim -17$~mag.  \cite{Galyam2007} and \cite{Galyam2009}
propose that the spectral properties
of SN~2005gl are best explained by heavy mass loss or mass ejections
closely correlated with the SN.  In fact, there is growing evidence
that pre-supernova bursts of mass loss, while not common, are also
not rare.   The most remarkable case is the eruption observed
two years prior to the peculiar Type~Ib SN~2006jc (\citealt{Pastorello2007}),
but the light curves of other SN show strong evidence for mass ejection
episodes shortly before the SN (e.g., SN~2006gy, SN~2005ap, SN~2006tf, SN~2007va,
see, e.g., \cite{Smith2008}, \cite{Kozlowski2010}).  The case of SN~2006jc may be particularly
apt since the spectral evidence for excess helium (\citealt{Branch1971})
suggests that SN~1961V was close to being a Type~Ib rather than a Type~IIn/pec.

Suppose we interpret SN~1961V in this context.  In this view, the
pre-SN light curve of SN~1961V is mapped into the pre-SN history
of mass loss, and these phases of mass loss are then mapped into
the post-SN light curve.  We modify the  \cite{Goodrich1989}
scenario as follows.  In quiescence, the progenitor is a hot star
with a low density fast wind. Sometime
before 1930 (likely closer to 1800), the star transitions
from the compact, hot ($T_* \sim 40000$~K) state with a low
density, high velocity wind to (on average) a cooler ($T_* \sim 7500$~K)
star with a high density, lower velocity wind.  Then, around
1955 (not 1960) the star undergoes an LBV (or other) eruption
to produce the pre-peak luminosity plateau, accompanied
by a further rise in the wind density and a higher wind
velocity.  The light curves in \cite{Branch1971} and \cite{Doggett1985}
appear to allow the outburst phase to commence earlier than 1960 due
to the poor quality of the magnitude limits from 1955 to 1960.
Then in December 1961, the star undergoes
core collapse and produces an SN, leading to the luminosity peak.  The
outgoing shock wave now interacts with the previous
mass loss history, where the first, more luminous
post-SN plateau is due to interactions with the
LBV eruption ejecta, and the second, longer plateau is due
to the wind emitted in the cool phase.  The luminosity
then drops dramatically when the shock wave reaches the low
density wind of the pre-explosion hot star phase.  

The observed velocities now represent the velocity of the expanding
shock wave rather than the wind.   Suppose we try to power the
light curve using the luminosity available from shock heating
the circumstellar medium (CSM),
\begin{equation}
   L = { v_s^3 \over 2 v_w } \epsilon \dot{M}
     \simeq  10^{7.7}\left({ \dot{M} \over 10^{-2} M_\odot/\hbox{year} }\right)
                 \left( { v_s \over 4000 \hbox{km/s} } \right)^3
                 \left( { 100 \hbox{km/s} \over v_w } \right)
                 \left( { \epsilon \over 0.1 } \right) L_\odot
\end{equation}
where $\dot{M}$ is the mass loss rate, $v_w$ is the wind velocity,
$v_s$ is the shock velocity and $\epsilon$ is the radiative efficiency (e.g. \citealt{Chugai1994}).
Here we have scaled $v_s$ to the $4000$~km/s suggested by the VLBI
observations of \cite{Chu2004}.  Using $2000$~km/s simply drives 
the required mass loss rates upwards while making it easier to
have extended, post-SN luminosity plateaus.

The difficult part about producing the first post-transient
luminosity plateau is its duration.  The length of the plateau $t_p$
should be roughly $t_p \simeq t_e v_w/v_s$, where $t_e$ is the
duration of the eruption prior to 1961.  If $t_e \simeq 1$~year
and $t_p \simeq 0.5$~years, as in the description of the light
curve by \cite{Goodrich1989}, the required velocity ratio $v_w/v_s \simeq 1/2$
seems unphysical.  However, if however, the eruption
commenced closer to 1955, so that $t_e \simeq 5$~years,
but was missed due to the shallowness of the observations
(\citealt{Branch1971}, \citealt{Doggett1985}), then we need
only have $v_s/v_s \simeq 1/10$.  Suppose we adopt
$v_s/v_w=5$, then reproducing the plateau luminosity of order
$L \simeq 2 \times 10^7 L_\odot$ ($m_{pg} \simeq 17$) requires
a mass loss rate of
\begin{equation}
      \dot{M} \simeq 0.03
                 \left( { 4000 \hbox{km/s} \over v_s } \right)^2
                 \left( { 5 v_w \over v_s  } \right)
                 \left( { 0.1 \over \epsilon } \right) M_\odot/\hbox{year}
\end{equation}
which is grossly consistent with an LBV eruption (\citealt{Humphreys1994}). This
rises to $10^{-1} M_\odot$/year if we use $v_s=2000$~km/s.  In either case, 
enough mass is involved that we may be also be underestimating the radiative 
efficiency (see \citealt{Smith2007}).

The shock then moves out through the lower density material from the
earlier ``S Doradus'' phase -- here we simply envision an extended
period where the star is on average producing a relatively dense
wind.  The necessary shock luminosity is now 10 times lower, and we are free
to make the ratio $v_s/v_w$ much larger since we have no definitive
time scale for the start of this phase beyond that it began before
$\sim 1930$.  Thus, the mass loss rate need only be
\begin{equation}
      \dot{M} \simeq 4 \times 10^{-4}
                 \left( { 4000 \hbox{km/s} \over v_s } \right)^3
                 \left( {  v_w \over 100\hbox{km/s} } \right)
                 \left( { 0.1 \over \epsilon } \right) M_\odot/\hbox{year},
\end{equation}
which is relatively easy for an LBV produce even outside of eruptions.
Here $v_s/v_w=40$, so the enhanced mass loss phase would have started
circa $1800$ in order to make $t_p \simeq 4$~years.  The biggest problem 
with this schematic is that the material ejected prior to the SN cannot 
itself form significant amounts of dust or the progenitor would have been 
self-obscured, similar to SN~2008S (see, e.g., \citealt{Prieto2008}).

Not only was the progenitor of SN~1961V massive, $M_{ZAMS} \gtorder 80 M_\odot$, but 
it must also have been relatively massive at death.  While appearing to be rich in helium 
(\citealt{Branch1971}), it was still a Type~II SN, rather than a Type~Ib or Ic.
We found no detailed
pre-supernova models for this mass and metallicity range, but it is in the 
regime that \cite{Heger2003} estimate would be weak Type~Ib/c fall-back SN
leading to black hole formation.  
The solar metallicity models in \cite{Woosley2002}
have already lost much of their helium to mass loss. For their low
metallicity models ($10^{-4}$ solar!), the star would need to be more massive 
at death than $M \simeq 30M_\odot$ in order to retain any hydrogen.  
If it was a fall-back SN forming a BH, SN~1961V was not notably sub-luminous.   
Alternatively, some massive stars may
still form NSs, as suggested by the existence of a magnetar in Westerlund~1.
The progenitor of this NS seems to require a $>40 M_\odot$ progenitor 
given the other massive stars in the cluster (\citealt{Muno2006}),  unless it
can be explained by binary evolution and mass transfer (\citealt{Belczynski2008}). Since it takes
virtually no mass to power accretion onto a $\sim 10 M_\odot$ black
hole at the Eddington limit compared to the ejected mass, we might expect
the newly formed BH to accrete at the Eddington limit for an extended 
period of time.  However, \cite{Perna2008} found no X-ray emission 
from the site to a limit of $L_X < 6 \times 10^{37}$~ergs/s 
(2-10~keV)\footnote{With a similar limit of $<2 \times 10^{37}$~ergs/s 
for the softer 0.3-8~keV band (Soria \& Perna 2010, private communication).}
in March 2000, corresponding to a limit of order 5\% of Eddington.

Finally, the existence of other Type~IIn supernova requiring major mass loss events 
shortly before collapse changes our prior on the likelihood of such correlations 
for SN~1961V.  Rather than being
bizarre, it is simply the closest example, one which is so close
that we could see the pre-SN activity.  SN~1961V, SN~2005gl and
their relatives are all cases where the correlated mass loss is
large and dramatic.  We usually assume that stars are evolving
quasi-statically in their last phases, with no indicators of
imminent death, yet this clearly does not hold for this class
of objects.  It is an interesting question whether this phenomenon
is limited to a special class of SN, as proposed by \cite{Galyam2007},
or that we presently only notice the most dramatic examples of a 
more ubiquitous phenomenon.  In either case, it appears that 
studies of SN progenitors should evolve from simple attempts
to obtain a single snapshot of the star to monitoring their 
behavior over their final years.  

\acknowledgements 

We would like to thank John Beacom, Jos\'e Prieto and Todd Thompson for discussions and comments, Rebecca
Stoll for estimating the local gas phase metallicity, and Robert Soria and Rosalba Perna
for providing a softer energy band X-ray flux limit.  CSK, DMS and KZS are supported by NSF grant AST-0908816.
This work is based in part on observations made with the Spitzer Space Telescope, which is operated by the 
Jet Propulsion Laboratory, California Institute of Technology under a contract with NASA, and
in part on observations made with the NASA/ESA Hubble Space Telescope, obtained from the data 
archive at the Space Telescope Institute. STScI is operated by the association of Universities for Research in Astronomy, Inc. 
under the NASA contract  NAS 5-26555. 
This research has made use of the NASA/IPAC Extragalactic Database (NED) which is operated by the Jet Propulsion Laboratory, 
California Institute of Technology, under contract with the National Aeronautics and Space Administration. 

{\it Facilities:}  \facility{Spitzer,HST}

\begin{deluxetable}{lllrrrrrr}
\tablecaption{Spitzer Observations of NGC~1058}
\tablewidth{0pt}
\tablehead{
\colhead{date} &\colhead{MJD} &\colhead{PI/Program} 
  &\colhead{$3.6\mu$m}
  &\colhead{$4.5\mu$m}
  &\colhead{$5.8\mu$m}
  &\colhead{$8.0\mu$m}
  &\colhead{$24\mu$m}
  &\colhead{$70\mu$m}
  }
\startdata
 2004-08-14 & 53231.31 &Fazio/69      &150   &150   &150   &150   &  0 &0 \\ 
 2004-08-25 & 53242.03 &Fazio/69      &  0   &  0   &  0   &  0   & 10 &9  \\
 2007-09-14 & 54357.66 &Kotak/40619   &300   &300   &300   &300   &  0 &0  \\
 2007-09-29 & 54372.22 &Kotak/40619   &  0   &  0   &  0   &  0   & 30 &0  \\
\enddata
\tablecomments{Exposure times are in seconds. The IRAC frame times were 30~s in
  both observations.  }
\label{tab:log}
\end{deluxetable}

\begin{deluxetable}{llrrrr}
\tablecaption{IRAC Photometry}
\tablewidth{0pt}
\tablehead{
\colhead{Source} &\colhead{Aperture} &\colhead{$[3.6]$} &\colhead{$[4.5]$} &\colhead{$[5.8]$} &\colhead{$[8.0]$}  \\
                 &                   &\colhead{(mJy)}   &\colhead{(mJy)}   &\colhead{(mJy)}   &\colhead{(mJy)}     }
\startdata
SN1961V area   & 2\farcs4   &  $0.0139 \pm 0.0011$  &  $0.0104 \pm 0.0014$  &  $0.0557 \pm 0.0062$  &  $0.1682 \pm 0.0096$  \\
               & 3\farcs6   &  $0.0226 \pm 0.0010$  &  $0.0160 \pm 0.0019$  &  $0.0690 \pm 0.0074$  &  $0.2070 \pm 0.0108$  \\
SN1961V/\#8    & DAOPHOT    &  $0.0078 \pm 0.0010$  &  $0.0079 \pm 0.0010$  &  $0.0288 \pm 0.0098$  &  $0.0614 \pm 0.0275$  \\
SN1961V/other  & DAOPHOT    &  $0.0084 \pm 0.0008$  &  $0.0081 \pm 0.0010$  &  $0.0096 \pm 0.0041$  &  $0.0571 \pm 0.0064$  \\
Star \#3       & 2\farcs4   &  $0.0071 \pm 0.0013$  &  $0.0031 \pm 0.0012$  &  $  < 0.019  $        &  $ < 0.028  $         \\
               & 3\farcs6   &  $0.0044 \pm 0.0021$  &  $0.0003 \pm 0.0017$  &  $0.0102 \pm 0.0084$  &  $0.0349 \pm 0.0103$  \\
               & DAOPHOT    &  $0.0103 \pm 0.0008$  &  $0.0073 \pm 0.0010$  &  \nodata               &  \nodata               \\
Star A         & 2\farcs4   &  $0.0190 \pm 0.0010$  &  $0.0115 \pm 0.0010$  &  $0.0346 \pm 0.0070$  &  $0.0238 \pm 0.0078$  \\
               & 3\farcs6   &  $0.0204 \pm 0.0017$  &  $0.0129 \pm 0.0016$  &  $0.0566 \pm 0.0075$  &  $0.0167 \pm 0.0080$  \\
               & DAOPHOT    &  $0.0256 \pm 0.0011$  &  $0.0151 \pm 0.0013$  &  \nodata               &  \nodata               \\
Star B         & 2\farcs4   &  $0.0147 \pm 0.0018$  &  $0.0075 \pm 0.0014$  &  $0.0209 \pm 0.0092$  &  $0.0529 \pm 0.0087$  \\
               & 3\farcs6   &  $0.0196 \pm 0.0029$  &  $0.0119 \pm 0.0019$  &  $0.0605 \pm 0.0081$  &  $0.0869 \pm 0.0088$  \\
               & DAOPHOT    &  $0.0149 \pm 0.0012$  &  $0.0094 \pm 0.0011$  &  \nodata               &  \nodata               \\
Star C         & 2\farcs4   &  $0.0064 \pm 0.0012$  &  $0.0057 \pm 0.0011$  &  $0.0367 \pm 0.0051$  &  $0.1031 \pm 0.0099$  \\
               & 3\farcs6   &  $0.0073 \pm 0.0019$  &  $0.0041 \pm 0.0017$  &  $0.0251 \pm 0.0063$  &  $0.1273 \pm 0.0101$  \\
               & DAOPHOT    &  $0.0056 \pm 0.0009$  &  $0.0062 \pm 0.0012$  &  \nodata               &  \nodata               \\
SN~2007gr      & 2\farcs4   &  $4.7079 \pm 0.0207$  &  $3.7626 \pm 0.0137$  &  $2.7023 \pm 0.0431$  &  $2.4228 \pm 0.1452$  \\
               & 3\farcs6   &  $5.1208 \pm 0.0313$  &  $4.0373 \pm 0.0192$  &  $3.0947 \pm 0.0225$  &  $3.4095 \pm 0.1182$  \\
               & DAOPHOT    &  $5.6602 \pm 0.1845$  &  $4.1676 \pm 0.3363$  &  \nodata               &  \nodata               \\
SN~1969L       & 2\farcs4   &  $0.0065 \pm 0.0013$  &  $ < 0.0060 $         &  $0.0093 \pm 0.0060$  &  $0.0165 \pm 0.0101$  \\
               & 3\farcs6   &  $0.0099 \pm 0.0012$  &  $ < 0.0065 $         &  $0.0208 \pm 0.0054$  &  $0.0013 \pm 0.0047$  \\
\enddata
\tablecomments{ Flux limits are $3\sigma$ limits. }
\label{tab:photometry1}
\end{deluxetable}

\begin{deluxetable}{lrr}
\tablecaption{MIPS Photometry}
\tablewidth{0pt}
\tablehead{
\colhead{Source} &\colhead{$[24]$} &\colhead{$[70]$}  \\
                 &\colhead{(mJy)}  &\colhead{(mJy)}   }
\startdata
SN1961V area   &  $0.226 \pm 0.0039$  &  $ < 8.0 $            \\
Star \#3       &  $0.089 \pm 0.0044$  &  \nodata                 \\
Star A         &  $0.043 \pm 0.0082$  &  \nodata                 \\
Star B         &  $0.094 \pm 0.0080$  &  $<190$ \\
Star C         &  $0.027 \pm 0.0049$  &  $<178$ \\
SN~2007gr      &  $3.155 \pm 0.0398$  &  \nodata \\
SN~1969L       &  $ < 0.063   $       &  \nodata                 \\
\enddata
\tablecomments{ Flux limits are $3\sigma$ limits. }
\label{tab:photometry2}
\end{deluxetable}

\end{document}